\documentclass[a4paper,11pt]{article}
\pdfoutput=1 

\usepackage{jcappub} 
\usepackage[T1]{fontenc} 
\usepackage{comment}
\usepackage{physics}
\usepackage{todonotes}
\usepackage{subcaption}
\usepackage{dsfont}
\usepackage[normalem]{ulem}
\usepackage{booktabs}
\usepackage{xcolor}
\usepackage{bm}

\title{\fontsize{22}{25}\selectfont\boldmath  Tunnelling out of Starobinsky inflation: \\
Raising the spectral tilt}

\author[a]{Jose A. R. Cembranos,}
\author[a]{Jesús Luque}
\author[a]{and Javier Rubio}
\affiliation[a]{Departamento de Física Teórica and Instituto de Física de Partículas y del Cosmos (IPARCOS-UCM), Universidad Complutense de Madrid, 28040 
Madrid, Spain}
\emailAdd{cembra@ucm.es}
\emailAdd{jesluque@ucm.es}
\emailAdd{javier.rubio@ucm.es}

\abstract{
We propose a hybrid realization of Starobinsky inflation in which the inflationary epoch ends through vacuum decay. The model consists of an effective two-field system with a metastable Starobinsky branch shifted with respect to the true one. During the observable stage, the inflaton slow-rolls along the false branch, until a first-order phase transition in an orthogonal direction connects it to the true branch and ends inflation abruptly. This old-inflation-like exit skips the last part of the would-be Starobinsky trajectory. As a result, the Cosmic Microwave Background pivot scale exits the Hubble radius further from the minimum  of the false branch than in ordinary Starobinsky inflation, raising the scalar spectral tilt $n_s$ while preserving the characteristic small  tensor-to-scalar ratio. This provides a simple way of moving Starobinsky inflation towards the high-\(n_s\) region favoured by recent ACT-related combinations. The same vacuum transition leaves a stochastic gravitational-wave relic whose peak frequency is controlled by the tunnelling timescale and the subsequent reheating history.
}

\begin{document}

\maketitle

\section{Introduction} 
\label{sec:Intro}

Inflation provides a compelling explanation for the large-scale homogeneity and spatial flatness of the Universe, as well as for the origin of the primordial density perturbations that seeded cosmic structure formation \cite{Guth:1980zm,Guth:1982pn,Linde:1981mu,Linde:1983gd,Linde:1993cn,Baumann:2009ds}. Historically, the first inflationary scenario was formulated as a period of false-vacuum expansion ending through a first-order phase transition (FOPT)~\cite{Guth:1980zm}. Its main difficulty was the graceful-exit problem: the exponential expansion of the false vacuum prevented the nucleated bubbles from percolating and completing the transition. This motivated the development of slow-roll inflation, in which accelerated expansion ends continuously as the inflaton approaches the minimum of its potential.

Among the simplest and most successful realizations of this paradigm is Starobinsky inflation \cite{Starobinsky:1980te}. Its plateau potential predicts a nearly scale-invariant scalar spectrum characterized by a tilt $n_s$, together with a small tensor-to-scalar ratio $r$, in agreement with the constraints inferred from the Planck and BICEP2/Keck data, $n_s=0.965\pm0.004$ and $r<0.036$ \cite{Planck:2018jri,BICEP:2021xfz,Tristram:2021tvh}. Recent ACT measurements, especially when combined with other cosmological datasets, have, however, renewed interest in scenarios predicting somewhat larger values of the scalar spectral index $n_s=0.974\pm0.003$ \cite{AtacamaCosmologyTelescope:2025nti,AtacamaCosmologyTelescope:2025blo,DESI:2024mwx,DESI:2024uvr}. Although the inferred value depends on the precise dataset combination, this preference places the standard Starobinsky prediction closer to the lower edge of the corresponding confidence regions, motivating both modifications of the original scenarios \cite{Drees:2025ngb,Zharov:2025zjg,Addazi:2025qra,Ketov:2025cqg, Cecchini:2024xoq} and critical assessments of the theoretical price of reconciling it with the preferred spectral tilt \cite{DelGrosso:2026zbg}. It is therefore interesting to ask whether the prediction for $n_s$ can be shifted while preserving the characteristic plateau dynamics and small tensor-to-scalar ratio of Starobinsky inflation.

FOPTs provide a natural mechanism for abruptly changing the cosmological evolution. Their semiclassical description involves the nucleation and expansion of true-vacuum bubbles \cite{Coleman:1977py,Callan:1977pt,Coleman:1980aw,Devoto:2022qen}, while completion of the transition requires sufficiently rapid nucleation, percolation, and bubble collisions. Such transitions may also produce stochastic gravitational waves and other nonequilibrium relics \cite{Caprini:2019egz,Caprini:2024hue}. FOPTs have long been invoked as a mechanism for terminating inflation, particularly in hybrid and false-vacuum scenarios \cite{Linde:1993cn,Copeland:1994vg,Cortes:2009ej,Ashoorioon:2015hya}. They have also been considered in open-inflation models, where the observable Universe is identified with the interior of a nucleated bubble \cite{Bucher:1994gb,Yamauchi:2011qq,Linde:1995xm}, and in scenarios where an inflation-triggered transition takes place in a spectator sector while inflation continues, including analytical and numerical studies of the resulting dynamics and gravitational-wave signatures \cite{Jiang:2015qor,Domenech:2018bnf,An:2022cce,Zou:2026wzi}. In these latter constructions, however, global completion of the transition is either unnecessary or the transition itself does not provide the direct exit from inflation. More broadly, curvature-induced transitions in spectator sectors during or immediately after inflation have been investigated analytically and through classical lattice simulations, including their implications for symmetry breaking, reheating, baryogenesis and gravitational-wave production \cite{Bettoni:2018pbl,Bettoni:2018utf,Bettoni:2019dcw,Bettoni:2021zhq,Kierkla:2023uzo,Laverda:2023uqv,Laverda:2024qjt,Bettoni:2024ixe,Goertz:2024gzw,Laverda:2025pmg,Rubio:2025egw,Laverda:2026slq}.

In this work, we construct an effective two-field realization of Starobinsky inflation in which the inflationary epoch ends through vacuum decay. The observable stage takes place along a metastable false-vacuum branch whose potential has the standard Starobinsky form but whose minimum is displaced relative to that of the true-vacuum branch. A FOPT in an approximately orthogonal field direction connects the two branches. Since the same global value of the inflaton corresponds to different displacements from their respective minima, a point that is still undergoing slow roll on the false branch can lie beyond the inflationary region of the true branch. The transition can therefore terminate inflation directly, realizing a hybrid first-order exit from the Starobinsky plateau. By truncating the final portion of the would-be Starobinsky trajectory, the transition alters the mapping between Cosmic Microwave Background (CMB) horizon exit and the end of inflation. For a fixed physical number of e-folds, the pivot scale therefore exits at a larger displacement from the false-branch minimum than in the standard scenario, shifting $n_s$ closer to scale invariance while further suppressing the already small tensor-to-scalar ratio.

A successful realization of this scenario must overcome the same obstacle that affected old inflation: nucleation alone does not ensure that the transition completes in a vacuum-dominated background. We therefore impose not only the conventional percolation threshold, but also the decrease of the physical false-vacuum volume and a sufficient bubble abundance for collisions. We construct an explicit benchmark satisfying these requirements and verify that the inflationary trajectory remains effectively single-field during the observable stage. We also examine the semiclassical consistency of the calculation, including the stability of both branches, the subhorizon character of the bubbles, and the suppression of Coleman--De~Luccia and Hawking--Moss gravitational effects.

Because the transition takes place at the end of inflation, the gravitational waves generated by the bubble dynamics are not subsequently inflated away. Their present-day spectrum nevertheless depends on the post-transition expansion history. A substantial fraction of the vacuum energy released by the transition is expected to be deposited in bubble walls, field gradients, and excitations of the symmetry-breaking sector \cite{Espinosa:2010hh,Wang:2022lyd,Cutting:2020nla}. If this energy subsequently forms a matter-like population, the fractional gravitational-wave density is diluted until radiation domination begins. Conversely, sufficiently rapid decay and thermalization of the symmetry-breaking sector lead to effectively instantaneous reheating and avoid this additional suppression. We retain a general dependence on the reheating duration and equation of state, while adopting instantaneous reheating for our numerical benchmark. For a transition at the Starobinsky scale, the resulting signal generically peaks at frequencies well above the bands of currently planned terrestrial and space-based interferometers \cite{Punturo:2010zz,ET:2025xjr,LISA,AEDGE:2019nxb}.

The paper is organized as follows. In Section~\ref{sec: a hybrid FOPT}, we introduce the effective two-field model and derive the conditions under which the FOPT ends inflation directly. Section~\ref{sec:inflationary_observables} determines the resulting inflationary predictions, including the self-consistent CMB matching and reheating history. In Section~\ref{sec:realizing_transition}, we compute the vacuum-decay rate and formulate the percolation and completion conditions during inflation, before presenting an explicit benchmark realization. Section~\ref{sec:GW_vacuum} estimates the resulting stochastic gravitational-wave signal. We summarize our conclusions and discuss possible extensions in Section~\ref{sec:discussion}. The appendices provide complementary Jordan-frame and $E$-model $\alpha$-attractor formulations and collect the main consistency checks of the effective description.

\section{Two-field model} 
\label{sec: a hybrid FOPT}

To describe a first-order exit from the Starobinsky plateau, we introduce an effective two-field setup containing the inflaton-scalaron $\phi$ and a symmetry-breaking field $\sigma$. The former controls the slow-roll evolution along a metastable branch, while the latter mediates the transition to the true vacuum. In the Einstein frame, the dynamics is governed by an effective action
\begin{equation}
S=\int d^4x\sqrt{-g}\left[\frac{M_P^2}{2}R-\frac12(\partial\phi)^2-\frac12(\partial\sigma)^2-U(\phi,\sigma)\right]\,.
\label{eq:Einstein_complete_action}
\end{equation}
Here, $M_P=2.44\times10^{18},\mathrm{GeV}$ is the reduced Planck mass and $R$ denotes the Ricci scalar. We decompose the scalar potential as
\begin{equation}
U(\phi,\sigma)=V_{\rm S}(\phi,\sigma)+V_\sigma(\sigma)\,,
\label{eq:complete_potential}
\end{equation}
where the effective Starobinsky sector is given by
\begin{equation}
V_{\rm S}(\phi,\sigma)=\frac34M_P^2M_\phi^2\left[1-{\cal A}(\sigma)\exp\left(-\beta\frac{\phi}{M_P}\right)\right]^2\,,
\label{eq:Starobinsky_sector}
\end{equation}
with $M_\phi$ setting the characteristic Starobinsky mass scale and $\beta=\sqrt{2/3}$. The potential for the symmetry-breaking field $\sigma$ is taken to be
\begin{equation}
V_\sigma(\sigma)=\Lambda^4\left(\frac{\sigma}{v_\sigma}\right)^2\left(\frac{\sigma}{v_\sigma}-1\right)^2\,,
\label{eq:sigma_potential}
\end{equation}
with constants $\Lambda$ and $v_\sigma$. Possible Jordan-frame and $\alpha$-attractor realizations of this effective structure are discussed in Appendices~\ref{sec:jordan_engineering} and \ref{sec:alpha_attractor_realization}.

The potential $V_\sigma$ has two degenerate minima at $\sigma=0$ and $\sigma=v_\sigma$. Through its coupling to the Starobinsky sector, the field $\sigma$ controls the position of the minimum along the $\phi$ direction, thereby generating two approximately parallel Starobinsky branches. We choose the coupling function to be
\begin{equation}
{\cal A}(\sigma)=1+\xi\frac{\sigma^2}{M_P^2}\,,
\label{eq:F_engineered}
\end{equation}
with $\xi\leq 0$ a dimensionless constant. At the two minima of $V_\sigma$, the coupling function takes the values
\begin{equation}
{\cal A}(0)=1\,,\qquad {\cal A}(v_\sigma)=1+q\,,\qquad q\equiv\xi\frac{v_\sigma^2}{M_P^2}\in(-1,0]\,.
\label{eq:F_engineered_endpoints}
\end{equation}
For fixed $\sigma$, the minimum of the Starobinsky sector along the $\phi$ direction is located at
\begin{equation}
\phi_0(\sigma)=\frac{M_P}{\beta}\ln {\cal A}(\sigma)\,.
\label{eq:phi0_sigma}
\end{equation}
Evaluating this expression at the two minima of $V_\sigma$ gives
\begin{equation}
\phi_0^{\rm TV}\equiv\phi_0(0)=0\,,\qquad \phi_0^{\rm FV}\equiv\phi_0(v_\sigma)=\frac{M_P}{\beta}\ln(1+q)<0\,.
\label{eq:branch_minima}
\end{equation}
At the corresponding points in the two-dimensional field space, both contributions to the potential vanish,
\begin{equation}
U\bigl(\phi_0^{\rm TV},0\bigr)=U\bigl(\phi_0^{\rm FV},v_\sigma\bigr)=0\,.
\label{eq:degenerate_minima}
\end{equation}
The trajectory $\sigma=0$ defines an exact stationary valley of the full potential. The second valley is continuously connected to $\sigma=v_\sigma$, although its coupling to the Starobinsky sector generally shifts it slightly away from this value. When the $\sigma$ direction is sufficiently heavy, this displacement is small, and the second valley is well approximated by $\sigma\simeq v_\sigma$. We adopt this approximation in the analytic discussion below.

For the non-trivial case $q<0$, one has ${\cal A}(v_\sigma)=1+q<{\cal A}(0)=1$. The branch connected to $\sigma=v_\sigma$ is therefore displaced towards negative values of $\phi$ relative to the exact valley at $\sigma=0$. Moreover, at fixed $\phi>0$, it lies at higher potential energy. We consequently identify the branch connected to $\sigma=v_\sigma$ as the false-vacuum branch and the $\sigma=0$ valley as the true-vacuum branch. Their horizontal separation is
\begin{equation}
\Delta\phi\equiv\phi_0^{\rm TV}-\phi_0^{\rm FV}=-\frac{M_P}{\beta}\ln(1+q)>0\,.
\label{eq:Deltaphi_engineered}
\end{equation}
Furthermore, the dependence of the Starobinsky sector on $\sigma$ can then be interpreted as a horizontal displacement along the $\phi$ direction through the identity
\begin{equation}
1-{\cal A}(\sigma)\exp\left(-\beta\frac{\phi}{M_P}\right)=1-\exp\left[-\beta\frac{\phi-\phi_0(\sigma)}{M_P}\right]\,.
\label{eq:shifted_factor}
\end{equation}
Along the exact true-vacuum valley, $\sigma=0$, one has $V_\sigma(0)=0$ and $\phi_0^{\rm TV}=0$, so the potential reduces to
\begin{equation}
U(\phi,0)=\frac34M_P^2M_\phi^2\left[1-\exp\left(-\beta\frac{\phi}{M_P}\right)\right]^2\,.
\label{eq:true_valley}
\end{equation}
Likewise, neglecting the small displacement of the false-vacuum valley from $\sigma=v_\sigma$, one has $V_\sigma(v_\sigma)=0$ and $\phi_0^{\rm FV}=-\Delta\phi$. The potential along this branch is therefore
\begin{equation}
U(\phi,v_\sigma)=\frac34M_P^2M_\phi^2\left[1-\exp\left(-\beta\frac{\phi+\Delta\phi}{M_P}\right)\right]^2\,.
\label{eq:false_valley}
\end{equation}
Within this approximation, the two branches have the same functional form and differ only by the horizontal displacement $\Delta\phi$. This equivalence becomes manifest by introducing a field variable adapted to each branch,
\begin{equation}
\varphi^{\rm FV/TV}\equiv\phi-\phi_0^{\rm FV/TV}\,,
\label{eq:branch_variables}
\end{equation}
in particular,
\begin{equation}
\varphi^{\rm TV}=\phi\,,\qquad \varphi^{\rm FV}=\phi+\Delta\phi\,.
\label{eq:branch_variables_explicit}
\end{equation}
In terms of the corresponding branch-adapted variables, the potential takes the standard Starobinsky form on either branch,
\begin{equation}
U_{\rm FV/TV}\bigl(\varphi^{\rm FV/TV}\bigr)=\frac34M_P^2M_\phi^2\left[1-\exp\left(-\beta\frac{\varphi^{\rm FV/TV}}{M_P}\right)\right]^2\,.
\label{eq:branch_potential}
\end{equation}
The field $\phi$ therefore serves as a global coordinate in the two-dimensional field space, whereas $\varphi^{\rm FV}$ and $\varphi^{\rm TV}$ measure the displacement from the minimum of the corresponding branch.

\begin{figure}
\centering
\includegraphics[scale=0.6]{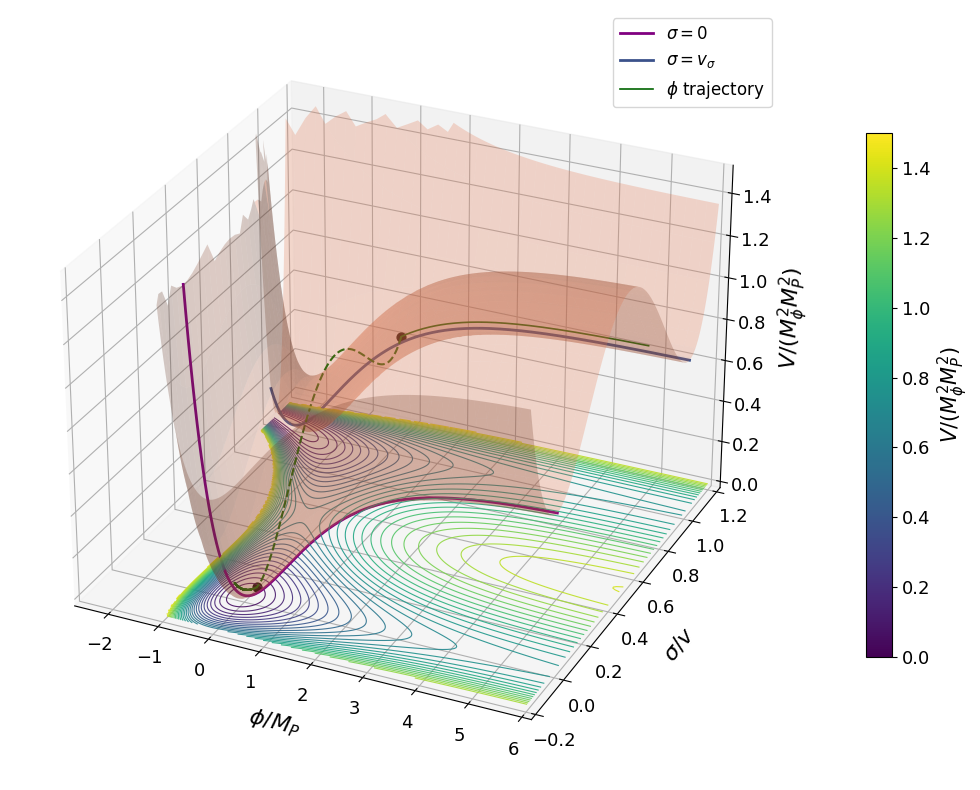}
\caption{Two-dimensional representation of the potential \eqref{eq:complete_potential} for $q=-0.81$. The purple and blue curves highlight the true- and false-vacuum branches, respectively. The green curve sketches the field-space trajectory: slow-roll evolution along the false-vacuum branch, followed by a FOPT in the $\sigma$ direction at $\phi_{*}=0.25M_P$ and subsequent evolution along the true-vacuum branch. The black markers indicate the configurations immediately before and after the transition.}
\label{Figure:potential}
\end{figure}

The resulting dynamics are illustrated in Figure~\ref{Figure:potential}. Inflation proceeds along the false-vacuum branch, whose minimum is shifted towards negative values of $\phi$. Since $\varphi^{\rm FV}=\varphi^{\rm TV}+\Delta\phi$, a fixed value of the global field $\phi$ can still lie on the plateau of the false branch while being much closer to the minimum of the true branch. The corresponding energy difference, together with the barrier separating the two valleys, allows the system to undergo a FOPT in the approximately orthogonal $\sigma$ direction. Because the transition takes place during inflation, and therefore before the formation of a thermal plasma, it proceeds through quantum tunnelling rather than thermal fluctuations.

To leading approximation, the transition changes the value of $\sigma$ while leaving the global field $\phi$ unchanged. The system, therefore, tunnels from the false-vacuum plateau to a lower-energy point on the standard Starobinsky branch. If the transition occurs at a value $\phi_{\rm *}<\phi_{\rm end}^{\rm TV}$, where $\phi_{\rm end}^{\rm TV}$ denotes the ordinary slow-roll endpoint of the true-vacuum branch, the transition terminates inflation directly. Otherwise, the system undergoes an additional period of slow-roll evolution after tunnelling.

When the transition provides the direct exit from inflation, the final part of the slow-roll trajectory that would otherwise be followed along the false branch is skipped. Consequently, modes leaving the horizon a fixed number of e-folds before the end of inflation do so at a larger displacement from the false-branch minimum than in standard Starobinsky inflation. Equivalently, the inflationary observables are evaluated at a larger effective number of Starobinsky e-folds, shifting the scalar spectral index $n_s$ towards larger values while slightly reducing the tensor-to-scalar ratio $r$. After the transition, the released vacuum energy is distributed among the residual scalaron configuration, bubble-wall motion, field gradients, and excitations of the symmetry-breaking sector, whose decay and thermalization determine the subsequent approach to radiation domination.

In the following sections, we analyze each stage of this mechanism in detail, from the conditions for a direct first-order exit to its consequences for inflationary observables, reheating, and gravitational-wave production.

\subsection{Background dynamics}
\label{sec:background_evolution}

To describe the cosmological evolution of the system, we consider a spatially flat Friedmann--Lemaître--Robertson--Walker background,
\begin{equation}
ds^2=-dt^2+a^2(t)\delta_{ij}dx^idx^j\,,
\label{eq:FLRW_background}
\end{equation}
where $a(t)$ is the scale factor and $H\equiv{\dot a}/{a}$ is the Hubble expansion rate. At the homogeneous level, the two scalar fields obey
\begin{equation}
\ddot\phi+3H\dot\phi+U_{,\phi}=0\,,\qquad \ddot\sigma+3H\dot\sigma+U_{,\sigma}=0\,.
\label{eq:background_field_equations}
\end{equation}
Their evolution is supplemented by the Friedmann equations,
\begin{equation}
3M_P^2H^2=\frac12\dot\phi^2+\frac12\dot\sigma^2+U(\phi,\sigma)\,,\qquad -2M_P^2\dot H=\dot\phi^2+\dot\sigma^2\,.
\label{eq:Friedmann_background}
\end{equation}
Before the phase transition, the system evolves approximately along the false-vacuum valley, $\sigma\simeq v_\sigma$. After tunnelling, it is transferred to the true-vacuum valley, $\sigma=0$. Equations~\eqref{eq:background_field_equations} and \eqref{eq:Friedmann_background} describe the homogeneous evolution along either branch. The additional contributions generated by bubble nucleation, field gradients and bubble-wall energy will be introduced when discussing the post-transition dynamics.

\subsection{Conditions for a direct first-order exit}
\label{sec:hybrid_exit}

We now determine the conditions under which the FOPT brings inflation directly to an end. Throughout this subsection, we work within the homogeneous slow-roll regime, approximate the false-vacuum valley by $\sigma\simeq v_\sigma$, and assume that tunnelling is both rapid compared to the Hubble timescale and approximately orthogonal to the $\phi$ direction. Under these assumptions, the global coordinate $\phi$ changes negligibly during the transition. The relations derived below therefore provide an analytic criterion for a direct first-order exit.

Since the two branches have the same functional form when expressed in terms of their respective branch-adapted variables, cf.~Eq.~\eqref{eq:branch_potential},  their ordinary slow-roll endpoint is determined by the same potential slow-roll parameter,
\begin{equation}
\epsilon_V\equiv\frac{M_P^2}{2}\left(\frac{V_{\rm S}'}{V_{\rm S}}\right)^2=\frac{4}{3\left(e^{\beta\varphi/M_P}-1\right)^2}\,,
\label{eq:epsilon_V_S}
\end{equation}
where the prime denotes differentiation with respect to the appropriate branch variable $\varphi$. The standard slow-roll estimate of the endpoint, $\epsilon_V=1$, gives
\begin{equation}
\varphi_{\rm end}\equiv\varphi_{\rm end}^{\rm TV}=\varphi_{\rm end}^{\rm FV}=\sqrt{\frac32}M_P\ln\left(1+\frac{2}{\sqrt3}\right)\simeq0.94\,M_P\,.
\label{eq:varphiend}
\end{equation}
Because the true-vacuum branch is unshifted, its endpoint in terms of the global coordinate is simply $\phi_{\rm end}^{\rm TV}=\varphi_{\rm end}$. The false-vacuum branch is instead displaced by $\phi_0^{\rm FV}=-\Delta\phi$, and hence
\begin{equation}
\phi_{\rm end}^{\rm FV}=\phi_0^{\rm FV}+\varphi_{\rm end}=-\Delta\phi+\varphi_{\rm end}\,.
\label{eq:false_end}
\end{equation}
Notice that the values of the field at which inflation would end, relative to the minimum of each branch, are the same ($\phi_{\rm end}$), as expected, since the potential is identical in both branches apart from the shift. Nonetheless, the global value, $\phi_{\rm end}$, is different in each branch precisely due to the shift, being that the reason we are noting $\phi_{\rm end}^{\rm FV}$, $\phi_{\rm end}^{\rm TV}$.

Let $\phi_\ast$ denote the value of the global inflaton coordinate when the phase transition becomes efficient. Immediately before tunnelling, the system is still inflating along the false-vacuum branch provided that
\begin{equation}
\varphi_\ast^{\rm FV}=\phi_\ast-\phi_0^{\rm FV}=\phi_\ast+\Delta\phi>\varphi_{\rm end}\,.
\label{eq:false_still_inflating}
\end{equation}
Because the transition is approximately orthogonal to the $\phi$ direction, the global coordinate remains nearly unchanged during tunnelling. The system therefore emerges on the true-vacuum branch at
\begin{equation}
\varphi_{\rm out}^{\rm TV}\simeq\varphi_\ast^{\rm TV}=\phi_\ast\,.
\label{eq:phi_out}
\end{equation}
Equivalently, the transition changes the branch-adapted coordinate according to
\begin{equation}
\varphi_\ast^{\rm TV}=\varphi_\ast^{\rm FV}-\Delta\phi\,.
\label{eq:branch_coordinate_transition}
\end{equation}
Equation~\eqref{eq:branch_coordinate_transition} provides the simplest geometric interpretation of the hybrid exit: tunnelling leaves the global coordinate $\phi$ approximately unchanged while reducing the displacement from the minimum of the relevant branch by an amount $\Delta\phi$, as illustrated in Figure~\ref{Figure:potential}.

Within the homogeneous slow-roll description, the transition ends inflation directly if the landing point lies outside the slow-roll region of the true-vacuum branch,
\begin{equation}
\varphi_\ast^{\rm TV}\lesssim\varphi_{\rm end}\,.
\label{eq:no_second_inflation}
\end{equation}
Using Eq.~\eqref{eq:branch_coordinate_transition}, this condition becomes
\begin{equation}
\Delta\phi\gtrsim\varphi_\ast^{\rm FV}-\varphi_{\rm end}\,.
\label{eq:Deltaphi_exit_condition}
\end{equation}
This is the central hybrid-exit condition: the separation between the two branches must be sufficiently large to transfer the system from an inflating point on the false-vacuum branch to a point outside the slow-roll region of the true-vacuum branch. Since the transition may also generate kinetic, gradient, and bubble-wall energy, Eq.~\eqref{eq:Deltaphi_exit_condition} should be regarded as a conservative homogeneous criterion.\footnote{If the condition in Eq.~\eqref{eq:no_second_inflation} is not satisfied, the system emerges inside the bubbles with $\varphi_\ast^{\rm TV}>\varphi_{\rm end}$ and undergoes an additional period of slow-roll inflation along the true-vacuum branch. Neglecting the kinetic and gradient energy generated during the transition, the duration of this second stage is estimated as
\begin{equation}
N_{\rm extra}=\frac{1}{M_P^2}\int_{\varphi_{\rm end}}^{\varphi_\ast^{\rm TV}}\frac{U_{\rm TV}}{U_{{\rm TV},\varphi}}\,d\varphi^{\rm TV}=\frac34\left[e^{\beta\varphi_\ast^{\rm TV}/M_P}-e^{\beta\varphi_{\rm end}/M_P}\right]-\frac{1}{2\beta M_P}\left(\varphi_\ast^{\rm TV}-\varphi_{\rm end}\right)\,.
\label{eq:Nextra_explicit}
\end{equation}
This is again a homogeneous slow-roll estimate. Any kinetic, gradient, or bubble-wall energy deposited during the transition would tend to shorten the additional inflationary stage.}

It is also useful to require that the transition does not place the field beyond the minimum, on the steep negative-$\phi$ side of the true-vacuum potential. Combining this requirement with the direct-exit condition gives the conservative landing window
\begin{equation}
0\lesssim\varphi_\ast^{\rm TV}=\phi_\ast\lesssim\varphi_{\rm end}\,.
\label{eq:no_overshoot}
\end{equation}
Using Eq.~\eqref{eq:branch_coordinate_transition}, the same window can be expressed in terms of the false-branch coordinate as
\begin{equation}
\varphi_\ast^{\rm FV}-\varphi_{\rm end}\lesssim\Delta\phi\lesssim\varphi_\ast^{\rm FV}\,.
\label{eq:no_overshoot_Deltaphi}
\end{equation}
The regime of primary interest is therefore the one in which the transition satisfies Eq.~\eqref{eq:no_overshoot_Deltaphi}, placing the system directly outside the slow-roll region of the true-vacuum branch without overshooting its minimum. In this case, inflation ends through the FOPT and the subsequent post-inflationary evolution begins immediately after tunnelling.

\section{Inflationary observables and cosmological matching}
\label{sec:inflationary_observables}

As discussed above, the observable stage of inflation takes place along the metastable false-vacuum branch, while the FOPT transfers the system to the true-vacuum branch and brings inflation to an end. Since the potential along the false-vacuum branch has the same functional form as the Starobinsky potential, the standard slow-roll relations remain unchanged. The difference lies instead in the endpoint of inflation: the transition occurs before the false branch reaches its ordinary slow-roll endpoint, thereby removing the final part of the would-be Starobinsky trajectory. This changes the relation between CMB horizon exit and the end of inflation, leading to modified predictions for the scalar spectral index $n_s$ and the tensor-to-scalar ratio $r$.

\subsection{False-branch inflation and skipped e-folds}
\label{sec:false_branch_inflation}

Within the approximation $\sigma\simeq v_\sigma$, the potential along the false-vacuum valley takes the standard Starobinsky form when expressed in terms of the corresponding branch-adapted field, cf.~Eq.~\eqref{eq:branch_potential}. At leading order in the slow-roll expansion, the amplitude of the scalar power spectrum, the scalar spectral index, and the tensor-to-scalar ratio are
\begin{equation}
A_s=\left.\frac{U_{\rm FV}}{24\pi^2M_P^4\epsilon_V}\right|_{\varphi_{\rm CMB}^{\rm FV}}\,,\qquad n_s=\left.1-6\epsilon_V+2\eta_V\right|_{\varphi_{\rm CMB}^{\rm FV}}\,,\qquad r=\left.16\epsilon_V\right|_{\varphi_{\rm CMB}^{\rm FV}}\,,
\label{eq:nsr_exact}
\end{equation}
with
\begin{equation}
\epsilon_V(\varphi)=\frac{4}{3\left(e^{\beta\varphi/M_P}-1\right)^2}\,,\qquad \eta_V(\varphi)=-\frac{4}{3}\frac{e^{\beta\varphi/M_P}-2}{\left(e^{\beta\varphi/M_P}-1\right)^2}\,,
\label{eq:slowroll}
\end{equation}
the potential slow-roll parameters. Here $\varphi_{\rm CMB}^{\rm FV}$ denotes the displacement from the minimum of the false-vacuum branch when the CMB pivot scale exits the Hubble radius. Thus, although the functional dependence of the observables is identical to that of standard Starobinsky inflation, their numerical values are modified through the determination of $\varphi_{\rm CMB}^{\rm FV}$.

The physical number of e-folds between CMB horizon exit and the onset of the FOPT is
\begin{equation}
N_{\rm CMB}=\frac{1}{M_P^2}\int_{\varphi_\ast^{\rm FV}}^{\varphi_{\rm CMB}^{\rm FV}}\frac{U_{\rm FV}}{U_{{\rm FV},\varphi}}\,d\varphi^{\rm FV}=\frac34\left[e^{\beta\varphi_{\rm CMB}^{\rm FV}/M_P}-e^{\beta\varphi_\ast^{\rm FV}/M_P}\right]-\frac{1}{2\beta M_P}\left(\varphi_{\rm CMB}^{\rm FV}-\varphi_\ast^{\rm FV}\right)\,.
\label{eq:NCMB_explicit}
\end{equation}
In ordinary Starobinsky inflation, the lower limit of this integral is the slow-roll endpoint $\varphi_{\rm end}$. In the present scenario, inflation instead terminates through the phase transition at $\varphi_\ast^{\rm FV}>\varphi_{\rm end}$. Consequently, for a fixed physical value of $N_{\rm CMB}$, Eq.~\eqref{eq:NCMB_explicit} places CMB horizon exit at a larger displacement from the false-branch minimum than in the standard scenario. Once $\varphi_{\rm CMB}^{\rm FV}$ has been determined, the inflationary observables follow directly from Eqs.~\eqref{eq:slowroll} and \eqref{eq:nsr_exact}.

The effect of the modified endpoint can be conveniently quantified by the number of would-be Starobinsky e-folds skipped as a result of the first-order exit,
\begin{equation}
\Delta N_{\rm skip}=\frac{1}{M_P^2}\int_{\varphi_{\rm end}}^{\varphi_\ast^{\rm FV}}\frac{U_{\rm FV}}{U_{{\rm FV},\varphi}}\,d\varphi^{\rm FV}=\frac34\left[e^{\beta\varphi_\ast^{\rm FV}/M_P}-e^{\beta\varphi_{\rm end}/M_P}\right]-\frac{1}{2\beta M_P}\left(\varphi_\ast^{\rm FV}-\varphi_{\rm end}\right)\,.
\label{eq:DeltaNskip_explicit}
\end{equation}
This motivates the definition of an effective number of Starobinsky e-folds,
\begin{equation}
N_{\rm eff}\equiv  N_{\rm CMB}+\Delta N_{\rm skip}\,.
\label{eq:Neff_def}
\end{equation}
This quantity represents the number of e-folds that would separate CMB horizon exit from the ordinary slow-roll endpoint if the phase transition were absent. At leading order in the large-$N$ expansion, the inflationary observables can therefore be written as \cite{Martin:2013tda}
\begin{equation}
n_s\simeq1-\frac{2}{|N_{\rm eff}|}\,,\qquad r\simeq\frac{12}{N_{\rm eff}^2}\,.
\label{eq:largeN_predictions}
\end{equation}
This approximation makes the physical effect of the mechanism particularly transparent. Since $\Delta N_{\rm skip}>0$, the first-order exit increases the effective number of Starobinsky e-folds associated with the CMB pivot scale. As a result, the scalar spectral index is shifted towards scale invariance, while the tensor-to-scalar ratio is reduced.

\subsection{CMB matching and post-transition reheating}
\label{sec:CMB_matching_reheating}

The effective number of e-folds $N_{\rm eff}$ introduced above captures the modification of the inflationary observables caused by the early termination of the false-vacuum trajectory. However, the physical number of e-folds $N_{\rm CMB}$ between CMB horizon exit and the end of inflation is not fixed independently of the subsequent evolution \cite{Barman:2025lvk}. It depends on the post-inflationary expansion history and, in particular, on the reheating duration $N_{\rm RH}$ and the average equation-of-state parameter $w_{\rm RH}$ during this period. A quantitative determination of $n_s$ and $r$ therefore requires the inflationary and reheating dynamics to be treated self-consistently.

For the pivot scale $k=0.002\,\mathrm{Mpc}^{-1}$, the standard matching relation between CMB horizon exit and the end of reheating can be written as \cite{Drees:2025ngb}
\begin{equation}
N_{\rm CMB}+\ln\left(\frac{\rho_\ast^{1/4}}{H_{\rm CMB}}\right)=64.82+\frac{3\bar w_{\rm RH}-1}{4}N_{\rm RH}\,,
\label{Eq: N_CMB prediction}
\end{equation}
where $\rho_\ast$ denotes the total energy density immediately before the phase transition and $H_{\rm CMB}$ is the Hubble rate when the CMB pivot scale exits the Hubble radius. Within the slow-roll approximation, these quantities are given by 
\begin{equation}
\rho_\ast\simeq U_\ast=\frac34M_P^2M_\phi^2\left(1-e^{-\beta\varphi_\ast^{\rm FV}/M_P}\right)^2\,,\qquad 
H_{\rm CMB}\simeq\frac{M_\phi}{2}\left(1-e^{-\beta\varphi_{\rm CMB}^{\rm FV}/M_P}\right)\,.
\label{Eq: H_CMB}
\end{equation}
Since the phase transition brings inflation to an end, we identify $\rho_{\rm end}\simeq\rho_\ast$. This assumes that tunnelling is rapid compared with the Hubble timescale.

After tunnelling, the system reaches the true-vacuum branch, $\sigma\simeq0$, where $\phi=\varphi^{\rm TV}$ according to Eq.~\eqref{eq:branch_variables_explicit}. The phase transition converts a substantial fraction of the false-vacuum energy into bubble-wall energy, field gradients, and excitations of the $\sigma$ sector. After bubble collisions and the subsequent relaxation of the inhomogeneous configuration, we assume that the dominant part of this energy is stored in oscillations of $\sigma$ around the true vacuum. Since the potential is approximately quadratic near $\sigma=0$,
\begin{equation}
V_\sigma(\sigma)\simeq\frac{1}{2}m_\sigma^2\sigma^2\,,\qquad m_\sigma^2=\frac{2\Lambda^4}{v_\sigma^2}\,, 
\end{equation}
the coarse-grained $\sigma$ component behaves on average as pressureless matter, with an effective equation-of-state parameter $\bar w_\sigma\simeq0$. We assume that reheating proceeds through the perturbative decay of this component into radiation, with an effective decay rate $\Gamma_\sigma$. Alternative mechanisms, including gravitational reheating supplemented by a suitable post-inflationary evolution, have also been considered in the literature \cite{Barman:2023opy}. Since the transition takes place in a vacuum-dominated background, the nucleated bubbles are expected to reach ultrarelativistic velocities, corresponding to the regime in which particle production associated with bubble expansion and collisions can be particularly efficient \cite{Watkins:1991zt,Falkowski:2012fb,Azatov:2020ufh,Azatov:2021ifm,Cembranos:2024pvy}. Neglecting the subleading contribution of the residual scalaron condensate, the corresponding energy-transfer equations are
\begin{equation} 
\dot\rho_\sigma+3H\rho_\sigma=-\Gamma_\sigma\rho_\sigma\,,\qquad\quad  \dot\rho_R+4H\rho_R=\Gamma_\sigma\rho_\sigma\,. \label{eq:rho_R_decay} 
\end{equation}
Radiation domination begins when the decay rate becomes comparable to the Hubble expansion rate, $H_{\rm RH}\simeq\Gamma_\sigma$. Assuming an approximately constant matter-like equation-of-state parameter during reheating, $w_{\rm RH} \simeq \bar w_{\sigma}\simeq0$, one obtains
\begin{equation}
T_{\rm RH}\simeq\left(\frac{90}{\pi^2g_{\ast,{\rm RH}}}\right)^{1/4}\sqrt{\Gamma_\sigma M_P}\,,\qquad N_{\rm RH}\simeq\frac{2}{3}\ln\left(\frac{H_\ast}{\Gamma_\sigma}\right)\,.
\label{eq:reheating_estimates}
\end{equation}
While keeping $N_{\rm RH}$ general in most of the following analysis, we will adopt the limiting case of effectively instantaneous reheating in the numerical benchmark considered in Section~\ref{sec: Benchmark}. This corresponds to assuming that the energy stored in the $\sigma$ sector is converted and thermalized on a timescale much shorter than a Hubble time, such that $\Gamma_\sigma\sim H_\ast$, $N_{\rm RH}\simeq0$ and $\rho_{\rm RH}\simeq\rho_\ast$. For $H_\ast\simeq5\times10^{-6}M_P$ and $g_\ast=106.75$, this gives $T_{\rm RH}^{\rm inst}\simeq 3\times10^{15}\,\mathrm{GeV}$.

\begin{figure}
\centering
\begin{subfigure}{0.48\textwidth}
\centering
\includegraphics[width=\linewidth]{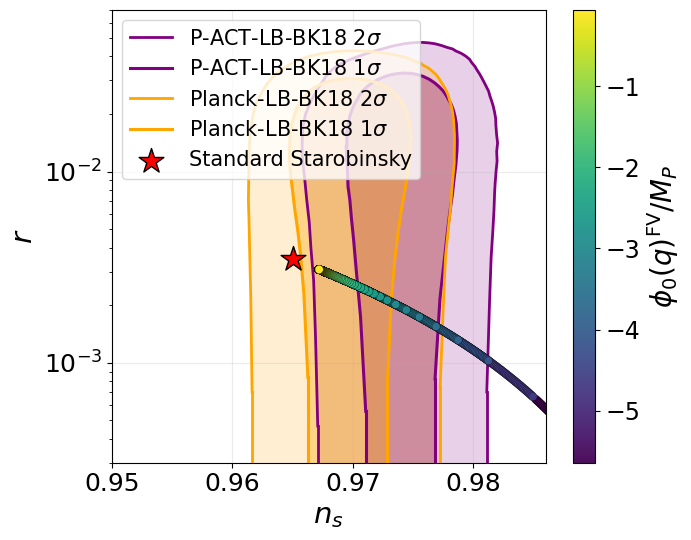}
\end{subfigure}
\hfill
\begin{subfigure}{0.48\textwidth}
\centering
\includegraphics[width=\linewidth]{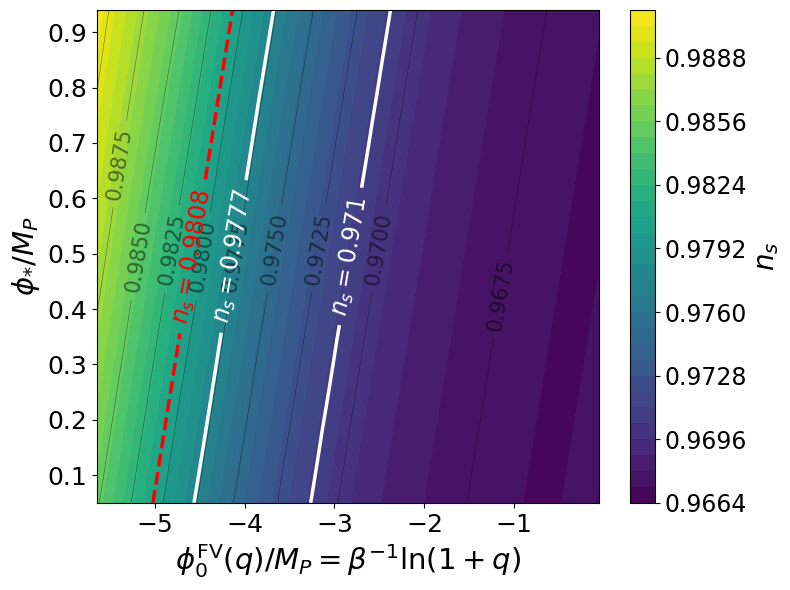}
\end{subfigure}
\caption{Inflationary predictions as functions of the model parameters $\phi_0^{\rm FV}$ and $\phi_\ast$. The orange and purple contours signal the  ACT and Planck $1\sigma$, $2\sigma$ confidence regions, respectively. The left panel shows the predictions in the $(n_s,r)$ plane as $\phi_0^{\rm FV}$ is varied, together with the observational confidence regions.  The star marks the standard Starobinsky prediction. The right panel shows the dependence of $n_s$ on both $\phi_0^{\rm FV}$ and $\phi_\ast$.  The solid white contours indicates the representative $1\sigma$ boundaries, while the dashed red contour indicates the upper $2\sigma$ bound for the ACT-Planck measurements \cite{AtacamaCosmologyTelescope:2025nti,AtacamaCosmologyTelescope:2025blo,DESI:2024mwx,DESI:2024uvr}}
\label{fig:inflation_predictions}
\end{figure}

The expressions in \eqref{eq:reheating_estimates} provide only an analytic estimate. The resulting reheating history could also have implications for the production of weakly coupled relics, as illustrated by ultraviolet freeze-in scenarios in Starobinsky inflation \cite{Bernal:2020qyu}. In the numerical analysis, the reheating evolution could be obtained by solving the coupled equations \eqref{eq:rho_R_decay} for a given value of $\Gamma_\sigma$. For each pair $(\phi_0^{\rm FV},\phi_\ast)$, the scalaron mass $M_\phi$ is fixed by the measured scalar amplitude, while the reheating equations would determine $N_{\rm RH}$ and $T_{\rm RH}$. In our case, since we are not specifying a particular reheating mechanism, we will assume instantaneous reheating, $N_{\rm RH}=0$. The matching relation in Eq.~\eqref{Eq: N_CMB prediction} and the e-fold relation in Eq.~\eqref{eq:NCMB_explicit} are then solved simultaneously for $N_{\rm CMB}$ and $\varphi_{\rm CMB}^{\rm FV}$, since $H_{\rm CMB}$ itself depends on $\varphi_{\rm CMB}^{\rm FV}$. Once this self-consistent solution is obtained, the corresponding values of $n_s$ and $r$ follow from Eqs.~\eqref{eq:slowroll} and \eqref{eq:nsr_exact}. The resulting inflationary predictions are displayed in Figure~\ref{fig:inflation_predictions}. The left panel shows how the prediction moves in the $(n_s,r)$ plane as the displacement of the false-vacuum branch is varied. The gap between Starobinsky’s typical result and the numerical results is due precisely to the assumption of instantaneous reheating. For a specific model with a given $N_{\rm RH}\neq0$, the mechanism presented here would also explore this region of parameter space. Even so, we see how the shift between branches allows us to modify the inflationary predictions, which is independent of the reheating model considered since it merely establishes the starting point. The right panel displays the combined dependence of $n_s$ on the branch displacement $\phi_0^{\rm FV}$ and the transition point $\phi_\ast$.

\section{Realizing the first-order phase transition}
\label{sec:realizing_transition}

A FOPT proceeds through the nucleation of true-vacuum bubbles within a metastable background. At zero temperature, the nucleation rate per unit physical spacetime volume is determined by the $O(4)$-symmetric Euclidean bounce \cite{Coleman:1977py,Coleman:1980aw,Athron:2023xlk}. Approximating the fluctuation prefactor by the characteristic inverse bubble volume, the decay rate is given by
\begin{equation}
\Gamma=R_0^{-4}e^{-S_4}\,,
\end{equation}
where $R_0$ is the initial bubble radius and $S_4$ is the Euclidean action evaluated on the bounce solution. Both quantities are computed numerically using the public code \texttt{AnyBubble} \cite{Masoumi:2016wot}.

The condition $\Gamma/H^4\gtrsim1$ indicates that bubble nucleation has become appreciable, corresponding approximately to one bubble nucleated per Hubble four-volume. It does not, however, guarantee that the transition percolates and completes. This distinction is particularly important in a vacuum-dominated background, where the exponential expansion may prevent the bubbles from filling space even when the nucleation rate is sizeable, as in the original old-inflation scenario \cite{Guth:1980zm,Guth:1982pn,Ellis:2018mja}.

\subsection{Percolation and completion during inflation}
\label{sec: percolation conditions}

The probability that a given point remains in the false vacuum at time $t$ is $P(t)=\exp[-f(t)]$, where
\begin{equation}
f(t)=\frac{4\pi}{3}\int_0^t\Gamma(t')a(t')^3r(t,t')^3dt'\,;\qquad r(t,t_i)=\int_{t_i}^t v_w\frac{d\bar t}{a(\bar t)}\,,
\label{eq: f(N) definition}
\end{equation}
and $r(t,t_i)$ is the comoving radius at time $t$ of a bubble nucleated at $t_i$, while $v_w$ is the bubble-wall velocity. The conventional percolation criterion, $P(t)\lesssim0.7$, corresponds approximately to $f(t)\gtrsim0.34$ and signals that a sizeable fraction of space has been converted to the true vacuum \cite{Ellis:2018mja}. During inflation, however, this condition alone is insufficient because the remaining false-vacuum regions continue to expand. Successful completion of the transition requires the physical false-vacuum volume, proportional to $a^3(t)P(t)=a^3(t)e^{-f(t)}$, to decrease. This leads to the stronger condition \cite{Ellis:2018mja,Zou:2026wzi}
\begin{equation}
\frac{df(t)}{dt}>3H\,;\qquad \frac{df(N)}{dN}>3\,,
\label{Eq: fraction of vacuum condition}
\end{equation}
where $N=\ln a$ increases over time.

In addition, a sufficiently large number of bubbles must nucleate within each physical Hubble volume for collisions to occur before the bubbles are separated by the accelerated expansion. Following Ref.~\cite{Zou:2026wzi}, we require
\begin{equation}
\frac{n_{\rm bub}}{H^3}\geq4\,;\qquad \frac{d}{dt}\left[n_{\rm bub}(t)a(t)^3\right]=\Gamma(t)P(t)a(t)^3\,,
\label{Eq: density condition}
\end{equation}
where $n_{\rm bub}$ denotes the physical number density of bubbles. We therefore identify the onset of the transition as the earliest time at which the percolation threshold, the decrease in the physical false-vacuum volume, and the bubble-abundance condition are simultaneously satisfied.

During the quasi-de Sitter stage, assuming slowly varying $H$ and relativistic bubble walls, $v_w\simeq1$, the relevant quantities can be expressed in terms of the number of e-folds as
\begin{equation}
\begin{aligned}\frac{n_{\rm bub}(N)}{H(N)^3}&=\int_{N_{\rm ini}}^{N}\left[\frac{H(N')}{H(N)}\right]^3\frac{e^{-S_4(N')-f(N')-3(N-N')}}{[H(N')R_0(N')]^4}\,dN'\,;\\ f(N)&=\frac{4\pi}{3}\int_{N_{\rm ini}}^{N}\frac{e^{-S_4(N')}}{[H(N')R_0(N')]^4}\left(1-e^{-(N-N')}\right)^3dN'\,.\end{aligned}
\label{Eq: f and nb e-folds}
\end{equation}
Here we have used the quasi-de Sitter expression for the comoving bubble radius,
\begin{equation}
r(t,t_i)\simeq\frac{1-e^{-(N-N_i)}}{a(t_i)H(N_i)}\,.
\end{equation}
In the present mechanism, $\phi$ rolls along the false-vacuum branch while $\sigma$ remains trapped near its metastable minimum. Since the barrier separating the two branches lies predominantly along the $\sigma$ direction, we evaluate the tunnelling solution in this direction at each background value of $\phi$. Using \texttt{AnyBubble}, we compute $S_4(\phi)$ and $R_0(\phi)$ and map the resulting decay rate onto the inflationary background. We then reconstruct $f(N)$, $df/dN$, and $n_{\rm bub}(N)/H^3$ to determine whether and when the transition completes.

\subsection{Benchmark realization}
\label{sec: Benchmark}

To illustrate the complete mechanism, we consider the potential in Eq.~\eqref{eq:complete_potential} with $\Lambda=5.5\times10^{-3}M_P$, $\xi=-10^4$, and $v_\sigma=9.695\times10^{-3}M_P$. These parameters give $q=-0.94$ and $\phi_0^{\rm FV}\simeq-3.45M_P$. The scalaron mass $M_\phi$ is determined self-consistently from the measured scalar amplitude. In particular, we obtain $M_\phi=9.57\times10^{-6}M_P$ versus the usual $M_\phi=1.23\times10^{-5} M_P$.

We first solve the homogeneous field equations \eqref{eq:background_field_equations}, together with the Friedmann equations \eqref{eq:Friedmann_background}, to determine the inflationary background. The transverse-mass hierarchy discussed in Appendix~\ref{app:EFT_control} ensures that isocurvature fluctuations are suppressed during the observable stage. As an additional numerical check, we also evaluate the turning rate of the two-field trajectory \cite{Kaiser:2015usz},
\begin{equation}
\frac{\omega}{H}=\frac{1}{H}\frac{\left|\dot{\phi}\,\ddot{\sigma}-\dot{\sigma}\,\ddot{\phi}\right|}{\dot{\phi}^{\,2}+\dot{\sigma}^{\,2}}\,.
\label{turn-rate}
\end{equation}
Throughout most of evolution, and in particular between approximately $45\text{--}70$ e-folds before the end of inflation, we find $\omega/H\sim10^{-5}$. Together with the hierarchy $m_{\sigma,{\rm FV}}^2\gg H^2$, this confirms that the observable evolution is accurately described by an effectively single-field trajectory.

\begin{figure}
\centering
\includegraphics[width=0.8\linewidth]{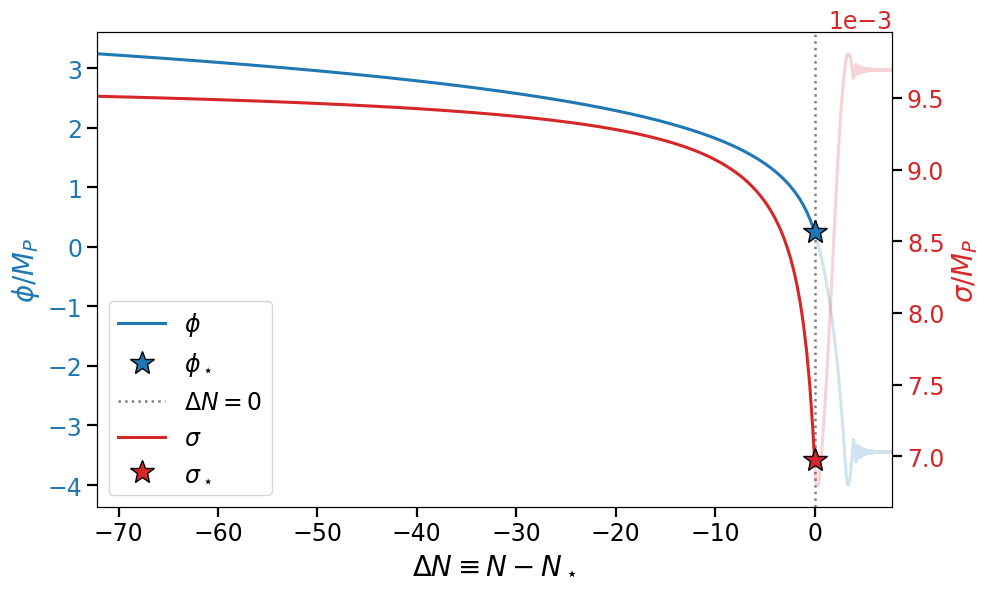}
\label{fig:grafica1}
\caption{Background evolution for the benchmark scenario. The plot displays the evolution of the fields during inflation. The stars and the vertical dotted line indicate the transition. Beyond this point, the faded curves show the hypothetical continuation of the homogeneous evolution in the absence of tunnelling. }
\label{fig:field evolution}
\end{figure}

\begin{figure}
\centering
\includegraphics[width=0.75\textwidth]{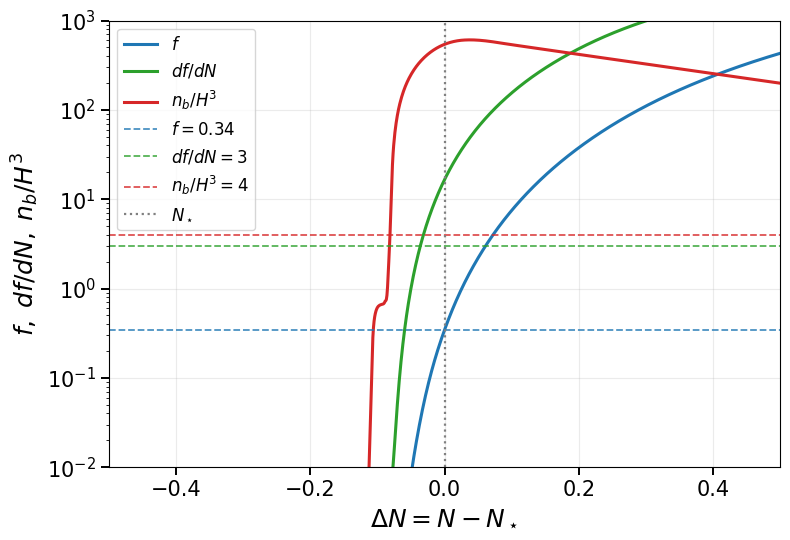}
\caption{Evolution of the false-vacuum exponent $f(N)$, its derivative $df/dN$, and the dimensionless bubble abundance $n_{\rm bub}/H^3$ around the transition time $N_\ast$. The reference lines indicate the thresholds $f=0.34$, $df/dN=3$, and $n_{\rm bub}/H^3=4$, while the vertical line marks the earliest time at which all three conditions are simultaneously satisfied.}
\label{Figure:percolation conditions}
\end{figure}

Figure~\ref{fig:field evolution} shows the evolution of the two fields along the false-vacuum branch. After determining this background trajectory, we compute the tunnelling rate at each value of $\phi$ using \texttt{AnyBubble}, reconstructing $f(N)$, $df/dN$, and $n_{\rm bub}/H^3$ from Eqs.~\eqref{eq: f(N) definition}--\eqref{Eq: f and nb e-folds}. We define the transition time $N_\ast$ as the earliest time at which $f(N_\ast)>0.34$, $df/dN|_{N_\ast}>3$, and $n_{\rm bub}(N_\ast)/H(N_\ast)^3>4$. For the benchmark parameters, these three conditions are first satisfied at $\phi_\ast\simeq0.25M_P$. Figure~\ref{Figure:percolation conditions} shows the evolution of the corresponding quantities near $N_\ast$, confirming that all three conditions remain satisfied after the transition begins.

Once $\phi_*$ has been determined, we assume that reheating occurs instantaneously due to mechanisms related to the phase transition, so that $N_{\rm RH}=0$ and $T_{\rm RH}=3\times10^{15}$ GeV. Note that for a specific mechanism with a certain effective $\Gamma_{\sigma}$, the reheating equations in \eqref{eq:rho_R_decay} would need to be solved, with the end of reheating defined by $\rho_\sigma=\rho_R$; this would also determine the reheating temperature 
\begin{equation}
T_{\rm RH}=\left.\left(\frac{30\rho_R}{\pi^2g_\ast}\right)^{1/4}\right|_{\rho_\sigma=\rho_R}\,.
\label{eq: reheating temperature}
\end{equation}

With $\phi_\ast$ and $N_{\rm RH}=0$ fixed, Eqs.~\eqref{Eq: N_CMB prediction}, \eqref{Eq: H_CMB}, and \eqref{eq:NCMB_explicit} are solved self-consistently to determine the CMB horizon-exit point. This gives $\phi_{\rm CMB}=2.20M_P$ (or equivalently $\varphi^{\rm FV}_{\rm CMB}=5.65 M_P$), $N_{\rm CMB}=58.53$, and $\Delta N_{\rm skip}=12.15$. The corresponding inflationary predictions, obtained from Eqs.~\eqref{eq:slowroll} and \eqref{eq:nsr_exact}, are $n_s=0.9725$ and $r=2.16\times10^{-3}$. This benchmark provides an explicit realization in which inflation proceeds along a shifted Starobinsky branch with $\phi_0^{\rm FV}\simeq-3.45M_P$ and ends through a FOPT at $\phi_\ast\simeq0.25M_P$. The removal of the final $\Delta N_{\rm skip}=12.15$ e-folds of the would-be Starobinsky trajectory shifts the inflationary predictions towards the high-$n_s$ region indicated by recent ACT observations while maintaining a small tensor-to-scalar ratio $r$.

Throughout this analysis, we treat the transition as a purely quantum vacuum decay and neglect both thermal and gravitational corrections \cite{Coleman:1980aw,Devoto:2022qen}. Thermal corrections are neglected because the transition occurs during inflation, before the formation of the thermal plasma associated with reheating. Gravitational corrections to the bounce are expected to be small when the initial bubble radius is much smaller than the Hubble radius, $R_0H_\ast\ll1$. For the benchmark point, we find $R_0H_\ast=5.9\times10^{-3}\ll1$, supporting the use of the flat-space $O(4)$ bounce approximation. We also verify in Appendix \ref{app:EFT_control} that neither Coleman–De Luccia nor Hawking–Moss gravitational effects modify our results in the parameter region considered.

\section{Gravitational waves from the vacuum transition}
\label{sec:GW_vacuum}

We now estimate the gravitational-wave signal generated by the FOPT. Since the transition occurs at the end of inflation, before the formation of a thermal plasma, we treat it as a vacuum transition. In this regime, there are no long-lived acoustic waves or magnetohydrodynamic turbulence. The dominant source is instead the inhomogeneous scalar-field configuration associated with the expanding bubble walls, their collisions, and their subsequent relaxation \cite{Caprini:2019egz,Caprini:2024hue,Cutting:2018tjt}.

The vacuum approximation requires any pre-existing radiation component to be negligible compared with the energy released during the transition, $\rho_R^\ast\ll\Delta U_\ast$. In the absence of particle production before the transition, this condition is automatically satisfied because any initial radiation component is exponentially diluted during inflation. The transition is therefore naturally vacuum-like in the scenario considered here. The relevant macroscopic parameters are the Hubble rate at the transition,
\begin{equation}
H_\ast^2\simeq\frac{\rho_\ast}{3M_P^2}\,,
\label{eq:Hstar_rhoS}
\end{equation}
the inverse duration of the transition,
\begin{equation}
\beta_{\rm PT}\equiv\left.\frac{d\ln\Gamma}{dt}\right|_{t_\ast}=-\left.\frac{dS_4}{dt}\right|_{t_\ast}-4\left.\frac{d\ln R_0}{dt}\right|_{t_\ast}\,,
\label{eq:beta_vac}
\end{equation}
or, equivalently,
\begin{equation}
\frac{\beta_{\rm PT}}{H_\ast}=\left.\frac{d\ln\Gamma}{dN}\right|_{N_\ast}=-\left.\frac{dS_4}{dN}\right|_{N_\ast}-4\left.\frac{d\ln R_0}{dN}\right|_{N_\ast}\,.
\label{eq:beta_vac_efolds}
\end{equation} 
In our numerical analysis, the contribution from the variation of the prefactor $R_0^{-4}$ is subleading compared with that arising from the variation of $S_4$. We therefore use the approximation
\begin{equation}
\beta_{\rm PT}\simeq-\left.\frac{dS_4}{dt}\right|_{t_\ast}\,,\qquad \frac{\beta_{\rm PT}}{H_\ast}\simeq-\left.\frac{dS_4}{dN}\right|_{N_\ast}\,.
\label{eq:beta_vac_approx}
\end{equation}
The remaining macroscopic parameter is the fraction of the total energy density released during the transition,
\begin{equation}
\alpha\equiv\frac{\Delta U_\ast}{\rho_\ast}\simeq\frac{\Delta U_\ast}{3M_P^2H_\ast^2}\,,
\label{eq:epsilon_star_def}
\end{equation}
where $\Delta U_\ast$ denotes the potential-energy difference between the two branches evaluated at $\phi_\ast$. Unlike the strength parameter commonly used for thermal transitions, which is normalized to the radiation density, the quantity $\alpha$ defined here is directly the released-energy fraction and therefore satisfies $0<\alpha\lesssim1$. Additionally, we define $\widetilde K=\kappa_\phi\alpha$, where $\kappa_{\phi}$ is the energy fraction transferred to the scalar-field source. For an efficient vacuum transition, most of the released energy is carried by scalar gradients and relativistic bubble walls, and we take $\kappa_\phi\simeq1$.

For an exponentially increasing nucleation rate, the characteristic physical bubble separation at collision is approximately \cite{Ellis:2018mja}
\begin{equation}
R_\ast\simeq(8\pi)^{1/3}\frac{v_w}{\beta_{\rm PT}}\,.
\label{eq:bubble_size}
\end{equation}
The applicability of a subhorizon treatment consequently requires
\begin{equation}
R_\ast H_\ast\simeq(8\pi)^{1/3}v_w\frac{H_\ast}{\beta_{\rm PT}}<1\,.
\label{eq:subhorizon_bubbles}
\end{equation}
We describe the present-day stochastic background using the broken-power-law template appropriate for strong transitions dominated by bubble collisions or highly relativistic shells \cite{Caprini:2024hue},
\begin{equation}
\Omega_{\rm GW}(f)=\Omega_p\frac{(n_1-n_2)^{(n_1-n_2)/a_1}}{\left[-n_2\left(\frac{f}{f_p}\right)^{-n_1a_1/(n_1-n_2)}+n_1\left(\frac{f}{f_p}\right)^{-n_2a_1/(n_1-n_2)}\right]^{(n_1-n_2)/a_1}}\,,
\label{GW spectra}
\end{equation}
where $f_p$ and $\Omega_p$ denote the present-day peak frequency and amplitude, respectively. We adopt $n_1=2.4$, $n_2=-2.4$, and $a_1=1.2$, as suggested by numerical simulations \cite{Caprini:2024hue, Cutting:2018tjt, Cutting:2020nla, Kosowsky:1992vn, Huber:2008hg, Weir:2016tov, Jinno:2019bxw}. 

The physical peak frequency at production is approximately $0.11\beta_{\rm PT}$. Its value today is therefore
\begin{equation}
f_p=0.11\beta_{\rm PT}\frac{a_\ast}{a_0}=0.11\beta_{\rm PT}e^{-N_{\rm RH}}\frac{T_0}{T_{\rm RH}}\left(\frac{g_{s,0}}{g_{s,\ast,{\rm RH}}}\right)^{1/3}\,.
\label{eq:f_peak_S}
\end{equation}
The peak amplitude produced by the scalar-field source can be parametrized as $\Omega_{p,\ast}=A_{\rm str}\widetilde K^2(H_\ast/\beta_{\rm PT})^2$, with $A_{\rm str}\simeq0.05$. Accounting for the subsequent cosmological evolution gives
\begin{equation}
h^2\Omega_p=1.64\times10^{-5}\left(\frac{100}{g_{\ast,{\rm RH}}}\right)^{1/3}e^{-N_{\rm RH}}A_{\rm str}\left(\frac{H_\ast}{\beta_{\rm PT}}\right)^2\widetilde K^2\,.
\label{eq: amplitude}
\end{equation}
The factor $e^{-N_{\rm RH}}$ appears because the gravitational-wave energy density redshifts as radiation, whereas the dominant oscillating $\sigma$ component behaves approximately as matter. Consequently, the fractional gravitational-wave density decreases as $a^{-1}$ during reheating.

For a matter-like reheating stage, the additional expansion between the transition and the onset of radiation domination is
\begin{equation}
e^{-N_{\rm RH}}=\frac{a_\ast}{a_{\rm RH}}=\left(\frac{\rho_{\rm RH}}{\rho_\ast}\right)^{1/3}=\left[\frac{\pi^2g_{\ast,{\rm RH}}T_{\rm RH}^4}{90M_P^2H_\ast^2}\right]^{1/3}\,.
\label{eq: redshift reheating}
\end{equation}
Taking $g_{*,\rm RH}\simeq g_{s*,\rm RH}\simeq 106.75$, we obtain 
\begin{equation}
f_p=5.41\times10^{9}\left(\frac{\beta_{\rm PT}/H_\ast}{100}\right)\left(\frac{H_\ast}{5\times10^{-6}M_P}\right)\left(\frac{3\times10^{15}\,{\rm GeV}}{T_{\rm RH}}\right)\left(\frac{100}{g_{s,\rm RH}}\right)^{1/3}e^{-N_{\rm RH}}{\rm Hz}\,.
\label{eq: final peak frequency}
\end{equation}
The present-day peak amplitude is correspondingly
\begin{equation}
\Omega_p h^2=8.22\times10^{-11}\widetilde K^2\left(\frac{100}{\beta_{\rm PT}/H_\ast}\right)^2\left(\frac{100}{g_{*,\rm RH}}\right)^{1/3}e^{-N_{\rm RH}}\,.
\label{Eq: final amplitude}
\end{equation}
This estimate leads to three main conclusions. First, because the phase transition terminates inflation, the generated signal is not diluted by any subsequent inflationary stage. Relative to the standard radiation-dominated transfer function, any additional suppression of the fractional gravitational-wave energy density arises solely from a matter-like reheating phase, if present. Second, for $H_\ast\sim10^{13},{\rm GeV}$, $T_{\rm RH}\sim10^{15},{\rm GeV}$, and $\beta_{\rm PT}/H_\ast\sim10\text{–}200$, the present-day peak frequency generically lies in the range $f_p\sim10^7\text{–}10^{10},{\rm Hz}$. Third, for $\alpha\sim0.3\text{–}1$ and $\kappa_\phi\simeq1$, the corresponding peak amplitude is typically $\Omega_ph^2\sim10^{-12}\text{–}10^{-9}$.~\footnote{Lower frequencies and larger amplitudes would require either an unusually slow transition or a more efficient scalar-field source. In the present mechanism, however, a sufficiently slow transition would prevent bubbles from colliding efficiently and would therefore jeopardize percolation.}

\begin{figure}
\centering
\includegraphics[width=0.75\textwidth]{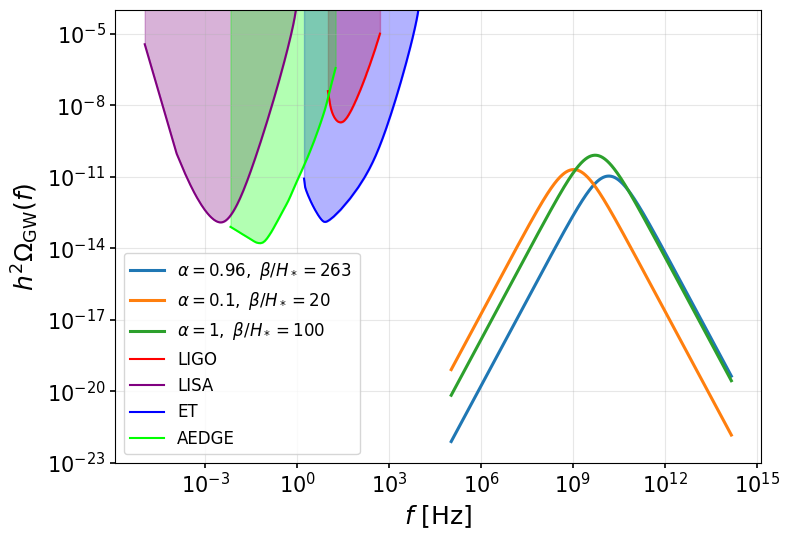}
\caption{Present-day gravitational-wave spectra for different values of $\alpha$, $\beta_{\rm PT}/H_\ast$, and $T_{\rm RH}$, with $H_\ast=5\times10^{-6}M_P$ fixed.}
\label{Figure: gravitational waves}
\end{figure}

These general estimates are illustrated in Figure~\ref{Figure: gravitational waves}, which shows representative spectra obtained by varying $\alpha$ and $\beta_{\rm PT}/H_\ast$ while fixing $H_\ast=5\times10^{-6}M_P$ and $T_{\rm RH}=3\times10^{15},{\rm GeV}$. The benchmark scenario of Section~\ref{sec: Benchmark}, shown in blue, has $\alpha=0.96$ and $\beta_{\rm PT}/H_\ast=263.4$. For $\kappa_\phi=1$, these values give approximately $f_p\simeq15,{\rm GHz}$ and $\Omega_p h^2\simeq2.8\times10^{-12}$, placing the signal well beyond the reach of planned terrestrial and space-based interferometers \cite{ET:2025xjr,LISA,AEDGE:2019nxb}. A matter-dominated reheating phase would shift the peak towards lower frequencies, but only at the cost of further suppressing its amplitude.

Note finally that, although the parameters governing the CMB shift also affect the gravitational-wave signal, the relation between the two observables is not one-to-one. The skipped inflationary interval is determined by the displacement between the two branches, controlled by $q$, together with the transition point $\phi_\ast$. These quantities also set the released-energy fraction $\alpha$ and, more weakly, the Hubble scale $H_\ast$. The gravitational-wave peak frequency and amplitude, however, depend sensitively on $\beta_{\rm PT}/H_\ast$, which is controlled by the variation of the bounce action and therefore carries additional information about the shape of the tunnelling barrier. A larger shift in $n_s$ may thus correlate with a larger released-energy fraction, while the detailed gravitational-wave spectrum remains primarily determined by the duration of the transition and the subsequent reheating history.

\section{Conclusions and outlook}
\label{sec:discussion}

In this work, we have presented a hybrid realization of Starobinsky inflation in which accelerated expansion ends through vacuum decay. During the observable stage, the system evolves along a metastable branch whose potential has the standard Starobinsky form when expressed in terms of the displacement from its own minimum. The FOPT then connects this trajectory to the true-vacuum branch while leaving the global inflaton coordinate approximately unchanged. Because the two branches have minima displaced relative to each other, a field value that is still slowly rolling on the false branch can correspond to a post-inflationary configuration on the true branch. The transition, therefore, modifies the endpoint of inflation without changing the local form of the potential responsible for generating the primordial perturbations.

The premature termination of the false-vacuum trajectory removes the final $\Delta N_{\rm skip}$ e-folds of the corresponding would-be Starobinsky evolution. Consequently, the CMB observables are determined by an effective number of Starobinsky e-folds, $N_{\rm eff}\simeq N_{\rm CMB}+\Delta N_{\rm skip}$. This shifts the scalar spectral index towards scale invariance while further reducing the already small tensor-to-scalar ratio, without introducing a qualitatively different inflationary energy scale. For the explicit benchmark considered here, the transition occurs at $\phi_\ast\simeq0.25M_P$ from a branch with $\phi_0^{\rm FV}\simeq-3.45M_P$, skipping approximately $\Delta N_{\rm skip}=12.15$ e-folds. After solving the CMB matching and reheating evolution self-consistently, we obtain $N_{\rm CMB}=58.53$, $n_s=0.9725$, and $r=2.16\times10^{-3}$. This illustrates how the mechanism can move the standard Starobinsky prediction into the representative high-$n_s$ confidence regions while preserving its characteristic small value of $r$.

A central requirement of our mechanism is that the transition actually completes in the vacuum-dominated background. We require simultaneously that the conventional percolation threshold be reached, that the physical false-vacuum volume decreases despite the accelerated expansion, and that the bubble abundance be sufficiently large for collisions. The benchmark satisfies all three conditions. We have also verified that the observable trajectory remains effectively single-field, that the relevant bubbles are subhorizon, and that neither the gravitational deformation of the localized bounce nor the competing Hawking--Moss channel modifies the decay process in the parameter region considered. The construction therefore provides an explicit realization of a first-order exit that avoids the graceful-exit problem of old inflation.

The transition also produces a stochastic gravitational-wave background. Since it occurs at the end of inflation, the generated signal is not erased by a subsequent inflationary stage. It is nevertheless diluted during the matter-like reheating period. For the benchmark parameters, $\alpha=0.96$, $\beta_{\rm PT}/H_\ast=263.4$, and $T_{\rm RH}=3\times10^{15}\,{\rm GeV}$, the spectrum peaks at approximately $f_p\simeq 15\,{\rm GHz}$ with amplitude $\Omega_p h^2\simeq2.8\times10^{-12}$. The signal is therefore well beyond the frequency bands and sensitivities of planned terrestrial and space-based interferometers \cite{ET:2025xjr,LISA,AEDGE:2019nxb}. More generally, the peak frequency and amplitude depend strongly on the duration of the transition and on the post-inflationary expansion history, while their relation to the shift in $n_s$ is indirect because the bounce action contains information about the barrier that is not fixed by the branch displacement alone.

Our treatment of the post-transition evolution has adopted a simple perturbative reheating framework as a reference scenario, assuming that the released energy is temporarily stored in a matter-like oscillating component before being transferred to radiation. The dynamics may be substantially richer \cite{Barman:2025lvk}. The released vacuum energy is, in reality, transferred directly into field gradients, bubble-wall motion, radiation, or particles produced during bubble expansion and collisions. A quantitative treatment of this regime requires following the nonlinear collision and relaxation of the bubbles, including backreaction and explicit couplings of $\sigma$ and $\phi$ to Standard Model particles, most likely through dedicated lattice simulations as in related studies of preheating dynamics \cite{Repond:2016sol}.

Such an analysis could modify the reheating history and the gravitational-wave transfer function, although the basic shift of the inflationary observables, which originates from the modified endpoint, would remain.

The proposed exit mechanism is not conceptually tied to the specific Starobinsky realization considered here. Its essential ingredients are simply a metastable inflationary branch and a lower-energy branch connected by vacuum decay; the detailed structure of the transition sector and of the true-vacuum branch is otherwise largely secondary, provided that the transition completes and leads to a viable post-inflationary evolution. The $E$-model embedding presented in Appendix~\ref{sec:alpha_attractor_realization} illustrates this broader applicability by extending the construction to general values of the attractor parameter $\bar\alpha$, thereby providing an additional handle on the tensor-to-scalar ratio. More generally, these observations motivate the exploration of analogous realizations in other plateau and attractor models, and suggest that vacuum decay may provide a robust and phenomenologically distinctive endpoint to inflation rather than merely an auxiliary event occurring during the inflationary era.

\acknowledgments

J.L acknowledges support from the Comunidad de Madrid under predoctoral contract PIPF-2024/TEC-34628. J.~R. is supported by a Ram\'on y Cajal contract of the Spanish Ministry of Science and Innovation with Ref.~RYC2020-028870-I. This research was further supported by the project PID2022-139841NB-I00 of MICIU/AEI/10.13039/501100011033 and FEDER, UE, and the 2025 Leonardo Grant for Scientific Research and Cultural Creation from the BBVA Foundation. The BBVA Foundation is not responsible for the opinions, comments, and content included in the project and/or its resulting outcomes, which are the sole and exclusive responsibility of the authors.

\appendix

\section{Jordan-frame realization}
\label{sec:jordan_engineering}

The effective two-field model introduced in the main text can be obtained from a scalar-tensor formulation of Starobinsky inflation in the Jordan frame. The key ingredient is a $\sigma$-dependent effective Planck mass, encoded in the function ${\cal A}(\sigma)$ defined in Eq.~\eqref{eq:F_engineered}. As $\sigma$ tunnels between its two vacua, ${\cal A}(\sigma)$ changes accordingly. After transforming to the Einstein frame, this change appears as the displacement of the Starobinsky minimum described in Section~\ref{sec:hybrid_exit}.

A convenient starting point is the auxiliary-field representation of the $R^2$ sector. Introducing an auxiliary scalar $\Phi$, we consider the Jordan-frame action
\begin{equation}
S_J=\int d^4x\sqrt{-g_J}\left[\frac12\Phi R_J-\frac{3M_\phi^2}{4}\left(\frac{\Phi}{M_P^2}-{\cal A}(\sigma)\right)^2-\frac12\frac{\Phi}{M_P^2}g_J^{\mu\nu}\partial_\mu\sigma\partial_\nu\sigma-\left(\frac{\Phi}{M_P^2}\right)^2V_\sigma(\sigma)\right]\,.
\label{eq:Jordan_engineered_action}
\end{equation}
Here, ${\cal A}(\sigma)$ and $V_\sigma(\sigma)$ are precisely the functions introduced in Eqs.~\eqref{eq:F_engineered} and \eqref{eq:sigma_potential}. The field $\Phi$ has no kinetic term in the Jordan frame and is therefore auxiliary. To clarify the gravitational origin of this construction, consider the first two terms in Eq.~\eqref{eq:Jordan_engineered_action}. In the absence of the explicit $\sigma$ kinetic and potential terms, eliminating $\Phi$ through its algebraic equation of motion gives
\begin{equation}
S_J\supset \frac{M_P^2}{2}\int d^4x\sqrt{-g_J}\left[{\cal A(\sigma)} R_J+\frac{R_J^2}{6M_\phi^2}\right]\,.
\label{eq:Jordan_R2_form}
\end{equation}
The function ${\cal A}(\sigma)$ therefore acts as a field-dependent effective Planck mass, while the second term is the usual Starobinsky $R_J^2$ correction. In particular, the transition from $\sigma\simeq v_\sigma$ to $\sigma\simeq0$ changes the coefficient of $R_J$ from $M_P^2(1+q)$ to $M_P^2$. 

We now perform the Weyl transformation and introduce the scalaron field through
\begin{equation}
g_{\mu\nu}=\frac{\Phi}{M_P^2}g_{\mu\nu}^J\,,\qquad \frac{\Phi}{M_P^2}=\exp\left(\beta\frac{\phi}{M_P}\right)\,,\qquad \beta=\sqrt{\frac23}\,.
\label{eq:Weyl_Phi}
\end{equation}
The corresponding metric relations are
\begin{equation}
\sqrt{-g_J}=\left(\frac{M_P^2}{\Phi}\right)^2\sqrt{-g}\,,\qquad g_J^{\mu\nu}=\frac{\Phi}{M_P^2}g^{\mu\nu}\,.
\label{eq:Weyl_metric_relations}
\end{equation}
Under this transformation, the non-minimal gravitational term generates the Einstein--Hilbert action together with a canonical kinetic term for $\phi$. The resulting action is
\begin{equation}
S=\int d^4x\sqrt{-g}\left[\frac{M_P^2}{2}R-\frac12(\partial\phi)^2-\frac12(\partial\sigma)^2-U(\phi,\sigma)\right]\,,
\label{eq:Einstein_engineered_action}
\end{equation}
which has exactly the form assumed in Eq.~\eqref{eq:Einstein_complete_action}. The Einstein-frame potential is indeed\footnote{The powers of $\Phi/M_P^2$ multiplying the last two terms in Eq.~\eqref{eq:Jordan_engineered_action} have been chosen so that the $\sigma$ sector acquires the desired form after the Weyl transformation. For the kinetic term, the factor $\Phi/M_P^2$ compensates the metric and volume factors in Eq.~\eqref{eq:Weyl_metric_relations}, leaving $\sigma$ canonically normalized. Similarly, the factor $(\Phi/M_P^2)^2$ multiplying $V_\sigma$ cancels the universal Weyl suppression of a Jordan-frame potential. Without this factor, a potential depending only on $\sigma$ in the Jordan frame would appear in the Einstein frame multiplied by $\exp(-2\beta\phi/M_P)$.}
\begin{equation}
U(\phi,\sigma)=\frac34M_P^2M_\phi^2\left[1-{\cal A}(\sigma)\frac{M_P^2}{\Phi(\phi)}\right]^2+V_\sigma(\sigma)=\frac34M_P^2M_\phi^2\left[1-{\cal A}(\sigma)\exp\left(-\beta\frac{\phi}{M_P}\right)\right]^2+V_\sigma(\sigma)\,.
\label{eq:Einstein_potential_from_Jordan}
\end{equation}
This reproduces precisely the decomposition in Eqs.~\eqref{eq:complete_potential} and \eqref{eq:Starobinsky_sector}. For fixed $\sigma$, the minimum of the Starobinsky contribution is determined by $\Phi(\phi_0)/M_P^2={\cal A}(\sigma)$ or, equivalently, 
\begin{equation}
\phi_0(\sigma)=\frac{M_P}{\beta}\ln{\cal A}(\sigma)\,,
\label{eq:Jordan_origin_phi0}
\end{equation}

\section{A shifted-vacuum $E$-model $\alpha$-attractor}
\label{sec:alpha_attractor_realization}

The mechanism developed in the main text admits a simple interpretation within the geometry of $E$-model $\alpha$-attractors \cite{Galante:2014ifa,Artymowski:2016pjz,Karananas:2016kyt}. Rather than modifying the asymptotic inflationary plateau, the tunnelling field dynamically shifts the location of the inflaton vacuum while leaving the infinite-distance boundary of field space unchanged. To avoid confusion with the strength parameter $\alpha$ used for the vacuum transition, we denote the attractor parameter by $\bar\alpha$. The Starobinsky realization studied throughout this work corresponds to $\bar\alpha=1$.

A convenient formulation of the $E$-model employs a positive dimensionless field $\rho$ with a second-order pole in its kinetic term,
\begin{equation}
S=\int d^4x\sqrt{-g}\left[\frac{M_P^2}{2}R-\frac{3\bar\alpha M_P^2}{4\rho^2}(\partial\rho)^2-\frac12(\partial\sigma)^2-U(\rho,\sigma)\right]\,.
\label{eq:alpha_attractor_pole_action}
\end{equation}
The inflationary boundary is located at $\rho=0$, which lies at an infinite geodesic distance in terms of the canonically normalized inflaton. A broad class of $E$-model potentials can be generated from a function $\mathcal{F}(\rho,\sigma)$ that remains regular at this boundary,
\begin{equation}
U(\rho,\sigma)=\mathcal{F}(\rho,\sigma)^2+V_\sigma(\sigma)\,.
\label{eq:alpha_attractor_general_potential}
\end{equation}
At the bosonic level, $\mathcal{F}$ acts as a generating function for the inflaton-dependent potential.

An $E$-model plateau requires $\mathcal{F}$ to approach a nonzero constant as $\rho\rightarrow0$. In addition, the mechanism considered in this work requires the location of the inflaton minimum to depend on $\sigma$. Denoting this position in the pole coordinate by $\rho_0(\sigma)$, the relevant conditions are
\begin{equation}
\mathcal{F}(0,\sigma)=\sqrt{\frac{3\bar\alpha}{4}}M_PM_\phi\,,\qquad \mathcal{F}\bigl(\rho_0(\sigma),\sigma\bigr)=0\,.
\label{eq:alpha_attractor_function_conditions}
\end{equation}
The standard $n=1$ $E$-model corresponds to a generating function whose zero is fixed at $\rho_0=1$. The present construction promotes this zero to a dynamical quantity controlled by the tunnelling field,
\begin{equation}
\rho_0(\sigma)=\frac{1}{\mathcal{A}(\sigma)}\,,
\label{eq:alpha_attractor_dynamic_zero}
\end{equation}
while preserving the plateau at the infinite-distance boundary $\rho=0$. The simplest regular generating function with these properties is linear in $\rho$,\footnote{More generally, replacing the bracket in Eq.~\eqref{eq:alpha_attractor_generating_function} by its $n$-th power generates the usual $E$-model potentials proportional to $\left(1-\mathcal{A}(\sigma)\rho\right)^{2n}$ \cite{Kallosh:2013yoa}.}
\begin{equation}
\mathcal{F}(\rho,\sigma)=\sqrt{\frac{3\bar\alpha}{4}}M_PM_\phi\left[1-\mathcal{A}(\sigma)\rho\right]\,.
\label{eq:alpha_attractor_generating_function}
\end{equation}
The resulting potential is therefore
\begin{equation}
U(\rho,\sigma)=\frac{3\bar\alpha}{4}M_P^2M_\phi^2\left[1-\mathcal{A}(\sigma)\rho\right]^2+V_\sigma(\sigma)\,.
\label{eq:alpha_attractor_pole_potential}
\end{equation}
The pole coordinate is related to the canonically normalized inflaton by
\begin{equation}
\rho=\exp\left(-\bar\beta\frac{\phi}{M_P}\right)\,,\qquad \bar\beta\equiv\sqrt{\frac{2}{3\bar\alpha}}\,.
\label{eq:alpha_attractor_canonical_map}
\end{equation}
Indeed,
\begin{equation}
\frac{3\bar\alpha M_P^2}{4\rho^2}(\partial\rho)^2=\frac12(\partial\phi)^2\,,
\label{eq:alpha_attractor_kinetic_map}
\end{equation}
and Eq.~\eqref{eq:alpha_attractor_pole_action} becomes
\begin{equation}
S=\int d^4x\sqrt{-g}\left[\frac{M_P^2}{2}R-\frac12(\partial\phi)^2-\frac12(\partial\sigma)^2-U(\phi,\sigma)\right]\,,
\label{eq:alpha_attractor_canonical_action}
\end{equation}
with
\begin{equation}
U(\phi,\sigma)=\frac{3\bar\alpha}{4}M_P^2M_\phi^2\left[1-\mathcal{A}(\sigma)\exp\left(-\bar\beta\frac{\phi}{M_P}\right)\right]^2+V_\sigma(\sigma)\,.
\label{eq:alpha_attractor_canonical_potential}
\end{equation}
For fixed $\sigma$, the minimum of the inflaton-dependent contribution is located at
\begin{equation}
\phi_0(\sigma)=\frac{M_P}{\bar\beta}\ln\mathcal{A}(\sigma)\,.
\label{eq:alpha_attractor_minimum}
\end{equation}
Introducing the displacement
\begin{equation}
\varphi=\phi-\phi_0(\sigma)\,,
\label{eq:alpha_attractor_displacement}
\end{equation}
the inflaton-dependent contribution to the potential takes the standard $n=1$ $E$-model form,
\begin{equation}
V_\varphi(\varphi)=\frac{3\bar\alpha}{4}M_P^2M_\phi^2\left[1-\exp\left(-\bar\beta\frac{\varphi}{M_P}\right)\right]^2\,.
\label{eq:alpha_attractor_standard_form}
\end{equation}
Thus, at fixed $\sigma$, the tunnelling field changes the position of the attractor minimum without modifying the shape of the potential measured relative to that minimum. The effective model studied in the main text is recovered by taking $\bar\alpha=1$ and $\bar\beta=\beta$. 

This construction also admits a natural embedding in the standard stabilizer-field formulation of $\mathcal N=1$ supergravity \cite{Kallosh:2011qk,Kallosh:2013yoa}. The hyperbolic Kähler geometry reproduces the kinetic pole, while suitable stabilizer couplings generate the scalar potential in Eq.~\eqref{eq:alpha_attractor_pole_potential}. In this realization, the tunnelling multiplet promotes the zero of the usual $E$-model stabilizer superpotential, and therefore the location of the corresponding bosonic minimum, to a field-dependent quantity, $T_0(\Sigma)={1}/{\mathcal A(\sqrt{2}\Sigma)}$,  while leaving the underlying hyperbolic geometry and inflationary attractor structure unchanged. The only modification with respect to the standard $E$-model construction is that the endpoint of the attractor trajectory becomes dynamical through its coupling to the tunnelling multiplet.

\section{Consistency conditions and regime of validity}
\label{app:EFT_control}

In this appendix, we collect the main consistency checks underlying the effective description used in the text. We first determine the characteristic inflationary scales, discussed the accuracy of the false-branch approximation, $\sigma\simeq v_\sigma$, verify that the evolution remains stable and effectively single-field in the transverse $\sigma$ direction and finally we confirm that there is no need to include gravitational corrections in the vacuum decay computations.

\begin{itemize}
\item{\bf Inflationary scales:} The measured amplitude of scalar perturbations fixes the scalaron mass $M_\phi$.  For $A_s=(2.1\pm0.03)\times10^{-9}$ \cite{Planck:2018jri} and the range $N_{\rm eff}\simeq55\text{--}84$ relevant to our analysis, Eqs.~\eqref{eq:branch_potential}, \eqref{eq:nsr_exact}, and \eqref{eq:slowroll} give $M_\phi\simeq(0.84\text{--}1.23)\times10^{-5}M_P$. Using the branch potential in Eq.~\eqref{eq:branch_potential} together with the Friedmann equation \eqref{eq:Friedmann_background}, the Hubble scale at the transition lies in the range $H_\ast\simeq(4.2\text{--}6.0)\times10^{-6}M_P\simeq(1.02\text{--}1.5)\times10^{13}\,{\rm GeV}$\,. The corresponding potential scale, obtained from Eq.~\eqref{Eq: H_CMB}, is $U_\ast^{1/4}\simeq(2.7\text{--}3.2)\times10^{-3}M_P\simeq(6.6\text{--}7.9)\times10^{15}\,{\rm GeV}$.
The first-order exit therefore modifies the endpoint of inflation and the predictions for $n_s$ and $r$ without introducing a qualitatively different inflationary energy scale. The characteristic values of $M_\phi$, $H_\ast$, and $U_\ast^{1/4}$ remain close to those of standard Starobinsky inflation.

\item {\bf Accuracy of the false-branch approximation:} The analytic discussion approximates the metastable valley by $\sigma=v_\sigma$. Its accuracy can be quantified by defining the exact false-vacuum trajectory through $U_{,\sigma}\bigl(\phi,\sigma_{\rm FV}(\phi)\bigr)=0$ with  $\sigma_{\rm FV}(\phi)=v_\sigma+\delta\sigma(\phi)$.  Expanding around $v_\sigma$ we have
\begin{equation}
\delta\sigma(\phi)\simeq-\frac{U_{,\sigma}(\phi,v_\sigma)}{U_{,\sigma\sigma}(\phi,v_\sigma)}
=-\frac{V_{S,\sigma}(\phi,v_\sigma)}{m_{\sigma,{\rm FV}}^2(\phi)}\,,
\label{eq:false_valley_shift}
\end{equation}
with $m_{\sigma,{\rm FV}}^2$ given in Eq.~\eqref{eq:msigmaf}. The relative displacement 
\begin{equation}
\frac{\delta\sigma(\phi)}{v_\sigma}\simeq
\frac{3\xi M_\phi^2e^{-\beta\phi/M_P}\left[1-(1+q)e^{-\beta\phi/M_P}\right]}
{m_{\sigma,{\rm FV}}^2(\phi)}
\label{eq:false_valley_relative_shift}
\end{equation}
therefore provides a direct estimate of the error associated with setting $\sigma=v_\sigma$. For the benchmark parameters, ${\delta\sigma(\phi)}/{v_\sigma}$ remains at the few-percent level during the observable stage and reaches only $\mathcal{O}(10^{-1})$ close to the transition. This behaviour is also visible in Figure~\ref{fig:field evolution}: the numerical trajectory closely follows $\sigma=v_\sigma$ over most of inflation and departs appreciably from it only as tunnelling becomes efficient. The approximation is therefore adequate for the analytic treatment, while the numerical background evolution follows the full two-field trajectory.

\item{\bf Transverse stability:} The observable inflationary stage is effectively single-field only if fluctuations in the transverse $\sigma$ direction are sufficiently heavy along the false-vacuum branch. Within the approximation $\sigma\simeq v_\sigma$, differentiating the potential in Eq.~\eqref{eq:complete_potential} gives
\begin{equation}
m_{\sigma,{\rm FV}}^2(\phi)=2\frac{\Lambda^4}{v_\sigma^2}+6M_\phi^2\frac{\xi^2v_\sigma^2}{M_P^2}e^{-2\beta\phi/M_P}-3\xi M_\phi^2e^{-\beta\phi/M_P}\left[1-(1+q)e^{-\beta\phi/M_P}\right]\,.
\label{eq:msigmaf}
\end{equation}
A controlled false-vacuum trajectory requires
\begin{equation}
m_{\sigma,{\rm FV}}^2(\phi_{\rm CMB})\gg H_{\rm CMB}^2\,,\qquad m_{\sigma,{\rm FV}}^2(\phi)>0\quad{\rm for}\quad\phi>\phi_\ast\,.
\label{eq:false_branch_conditions}
\end{equation}
The first condition suppresses isocurvature fluctuations during the observable stage, while the second guarantees that the false-vacuum branch remains locally stable until tunnelling becomes efficient.

For $\xi<0$, $q\in(-1,0]$, and $\phi>0$, all terms in Eq.~\eqref{eq:msigmaf} are non-negative. Regarding the presence of a barrier between the two branches, a simple sufficient condition is obtained by requiring its height $V_\sigma\left({v_\sigma}/{2}\right)$ at $\sigma=v_\sigma/2$ to exceed the asymptotic Starobinsky plateau $U_{\rm FV}(\varphi^{\rm FV}\rightarrow\infty)$. Using Eqs.~\eqref{eq:sigma_potential} and \eqref{eq:branch_potential}, this gives $\Lambda^4\gtrsim12M_\phi^2M_P^2$. Together with $H^2\lesssim M_\phi^2/4$, this implies
\begin{equation}
\frac{m_{\sigma,{\rm FV}}^2}{H^2}\gtrsim96\frac{M_P^2}{v_\sigma^2}\,.
\label{eq:false_branch_hierarchy}
\end{equation}
The transverse mode is therefore safely heavier than the Hubble scale for the sub-Planckian values of $v_\sigma$ considered here, and the observable evolution is effectively single-field.

After tunnelling, the system evolves along the exact true-vacuum valley at $\sigma=0$. The corresponding transverse mass is
\begin{equation}
m_{\sigma,{\rm TV}}^2(\phi)=\frac{2\Lambda^4}{v_\sigma^2}-3\xi M_\phi^2e^{-\beta\phi/M_P}\left[1-e^{-\beta\phi/M_P}\right]\,.
\label{eq:msigmat}
\end{equation}
For $\xi<0$ and $\phi>0$, the second term is positive. At the true-vacuum minimum, $\phi=0$, the Starobinsky contribution vanishes and the transverse mass is entirely determined by the barrier term. The same hierarchy that keeps the transverse mode heavy along the false-vacuum branch therefore guarantees that the system remains stable and confined to the true-vacuum valley after the transition.

\item \textbf{Gravitational corrections to vacuum decay:} Since the transition occurs during inflation, we must verify both that the de Sitter background does not significantly deform the localized bounce and that the competing Hawking--Moss channel is negligible. Gravitational corrections to the localized Coleman--De~Luccia bounce are suppressed when the bubble is much smaller than the Hubble radius, $R_0H_\ast\ll1$. 
For the benchmark scenario, $R_0H_\ast=5.9\times10^{-3}$, showing that the nucleated bubbles are deeply subhorizon and justifying the use of the flat-space $O(4)$ bounce computed with \texttt{AnyBubble}.

A distinct possibility in de Sitter space is the Hawking--Moss transition, which corresponds to the homogeneous displacement of an entire Hubble patch to the top of the barrier rather than to the nucleation of a localized bubble \cite{Hawking:1981fz}. Holding $\phi$ fixed at its value during the transition, the corresponding decay exponent is
\begin{equation}
B_{\rm HM}=24\pi^2M_P^4\left(\frac{1}{U_{\rm FV}}-\frac{1}{U_{\rm top}}\right)\,,
\label{eq:BHM}
\end{equation}
where $U_{\rm FV}=U(\phi_\ast,\sigma_{\rm FV})$ and $U_{\rm top}=U(\phi_\ast,\sigma_{\rm top})$. For a barrier height $\Delta U_b\equiv U_{\rm top}-U_{\rm FV}\ll U_{\rm FV}$, this reduces to
\begin{equation}
B_{\rm HM}\simeq\frac{8\pi^2}{3H_\ast^4}\Delta U_b\,.
\label{eq:BHM_approx}
\end{equation}

The Hawking--Moss solution always exists formally, but it dominates the decay only when the barrier is sufficiently broad compared to the de Sitter curvature scale. Conversely, if the barrier is sufficiently steep, the decay proceeds through the nucleation of localized Coleman--De~Luccia bubbles. A sufficient condition for the existence of such a localized bounce is ${|\lambda_-^{\rm top}|}/{H_{\rm top}^2}>4$. 
where $\lambda_-^{\rm top}$ is the negative eigenvalue of the field-space Hessian at the top of the barrier and $H_{\rm top}^2=U_{\rm top}/(3M_P^2)$. In the parameter region considered here, the unstable direction is approximately aligned with $\sigma$, so that $\lambda_-^{\rm top}\simeq U_{\sigma\sigma}^{\rm top}$. For the barrier potential in Eq.~\eqref{eq:sigma_potential}, we have
\begin{equation}
V_b=\frac{\Lambda^4}{16}\,,\qquad
|V_{\sigma\sigma}^{\rm top}|=\frac{\Lambda^4}{v_\sigma^2}=\frac{16V_b}{v_\sigma^2}\,,
\label{eq:barrier_top_curvature}
\end{equation}
and the steep-barrier condition becomes
\begin{equation}
V_b>\frac14v_\sigma^2H_{\rm top}^2\,.
\label{eq:Vb_HM_bound}
\end{equation}

For the benchmark parameters, $|U_{\sigma\sigma}^{\rm top}|/H_{\rm top}^2\sim\mathcal{O}(10^5)\gg4$, while Eq.~\eqref{eq:BHM} gives $B_{\rm HM}\sim\mathcal{O}(10^{12})$, far larger than the numerically computed localized-bounce action $S_4$. Vacuum decay is therefore exponentially dominated by localized bubble nucleation. The combined hierarchies
\begin{equation}
R_0H_\ast\ll1\,,\qquad
\frac{|U_{\sigma\sigma}^{\rm top}|}{H_{\rm top}^2}\gg4\,,\qquad
B_{\rm HM}\gg S_4\,,
\end{equation}
show that neither gravitational corrections to the localized bounce nor Hawking--Moss transitions modify the dynamics in the parameter region considered.

\end{itemize}

\bibliographystyle{JHEP}
\bibliography{biblio}

\end{document}